# Evolution of exploding granules from coordinated observations by THEMIS, IRIS, SDO/HMI, and HINODE, and a simulation

T. Roudier [1]      J.M. Malherbe[2]      B. Gelly [3]      R. Douet [3]
Z. Frank[4]      K. Dalmasse [1]

July 24, 2020

[1] Institut de Recherche en Astrophysique et Planétologie, Université de Toulouse, CNRS, UPS, CNES 14 avenue Edouard Belin, 31400 Toulouse, France
[2] Observatoire de Paris, LESIA, 5 place Janssen, 92195 Meudon, France, PSL Research University, CNRS, Sorbonne Universités, Sorbonne Paris Cité [3] THEMIS, INSU,CNRS,UPS 3718 IAC, Via Lactea , ES-38200 La Laguna, Tenerife - Spain [4] Lockheed Martin Solar and Astrophysics Laboratory, Palo Alto, 3251 Hanover Street, CA 94303, USA
thierry.roudier@irap.omp.eu

## Abstract

Exploding granules constitute the strongest horizontal flows on the quiet Sun and contribute to the structure of the surface horizontal velocity fields which build the large-scale organization of the discrete magnetic field.In this work we explore exploding granule expansion through the observations of the ground-based THEMIS telescope, IRIS, SDO, and the Hinode space-borne instruments, and finally with the magnetohydrodynamics simulation. We evaluate the detection and the expansion of exploding granules at several wavelengths and at various spatial and temporal resolutions.To analyze the different temporal sequences, two methods of image segmentation are applied to select the granules. The first allows us to follow individually the exploding granules observed simultaneously by THEMIS, IRIS, and SDO. The second uses long time independent sequences from THEMIS, IRIS, SDO, Hinode, and a simulation. In the first method (called manual) the segmentation isolates the cell of the granules (bright granules and intergranular parts), while in the second method (called statistical) only the bright part of the granules are isolated. The results obtained with simultaneous or distinct temporal observations using the two methods of segmentation are in good agreement. The granule area evolves linearly with an expansion velocity that decreases with the radius. A rapid decrease in the velocity expansion in the first two minutes is observed. The detection and measurement of the dynamics of the explosive granules can be performed from ground- and space-based instruments. Our work reveals the usefulness of SDO data, with low spatial resolution, to study the dynamics of the exploding granules all over the solar surface.

# 1 Introduction

The whole solar surface, covered with convective cells, is renewed every 10 to 15 minutes. These convective cells are known as solar granulation. Of the different types of granules the most dynamical phenomena are explosive granules (also called exploders). These granules push their surrounding vigorously and disintegrate into smaller fragments [Oba, 2018, Rieutord et al., 2001, Hirzberger et al., 1999, Title et al., 1986]. These granules represents only 2.5% of the total surface [Namba, 1986, Title et al., 1986], but have a large influence on the dynamics of the photosphere. For example, the surface properties of the supergranules may be explained as nonlinear interaction between granules triggered by exploding granules [Rieutord et al., 2001]. Recent works also show their involvement in the generation and diffusion of the magnetic field in the quiet Sun as well as their contribution to the formation of the photospheric network [Malherbe et al., 2018, Roudier et al., 2016, 2003]. Palacios et al. [2012], Moreno-Insertis et al. [2018], Zhang et al. [2009], and De Pontieu [2002] show the link between exploders and magnetic concentration, the magnetic field emergence in mesogranular size exploding granules observed with sunrise IMaX. The contribution of exploding granules in the magnetic field amplification is described in Rempel [2018]. The motions at the periphery of the pores are substantially and continuously influenced by the external plasma flows deposited by the explosive events [Sobotka, 2003, Roudier et al., 2002]

From single spectral lines compared to the time–space diagrams, it has been shown that the exploding granules amplify the acoustic oscillation [Ellwarth, 2018, Rast et al., 1993]. Larger granules amplify these oscillations stronger than smaller ones [Ellwarth, 2018]. The evolution and temporal height properties of the exploders has been studied by different authors [Ellwarth, 2018, Oba, 2018, Oba et al., 2017, Berrilli et al., 2002, Roudier et al., 2001]. A relation is observed between the exploding granule intensity located at the bottom of the photosphere and the bright plumes located at the top of the photosphere in the downflow. These brightnesses appear around the granule where the downward velocities are the most intense [Roudier et al., 2001]. Berrilli et al. [2002] showed a deceleration in the center of an upward flow before the appearance of dark areas, which suggests a decrease in the vertical velocity before the plume formation. They also observed a downflow ring around the new granules at the end of the fragmentation process. Recently, Fischer et al. [2017] described the chromospheric impact of an exploding solar granule. They showed that exploding granules can trigger upward-propagating shock fronts that dissipate in the chromosphere. Exploding granules have also been confirmed from theory and simulations [Stein and Nordlund, 1998, Rast, 1995, Rast and Toomre, 1993, Spruit and Title, 1990]. A detailed bibliography on exploding granules can be found in recent works [Ellwarth, 2018, Oba, 2018].

All these studies focus on the importance of the exploders on the dynamics of the photosphere and their influence in the evolution of the discrete magnetic field forming the network of the quiet Sun. Today, we have access to various ground-based and space-borne instruments, allowing us to study these explosive granules at different wavelengths and various spatial resolutions.

To date, the description of the exploding granule is mainly based on the disk center observations at high spatial resolution. As such energetic convective elements clearly influence the properties of the photosphere and also the evolution of magnetic field in the quiet Sun (and probably close to active regions), our main scientific goal is to detect their density and analyze their physical characteristics all over the Sun. The first step of our project, described in this paper, is to compare the exploding granule detection and properties from SDO (space-based) full Sun images (moderate spatial resolution) and high spatial resolution observations from ground- and space-based instruments at the disk center. The second step is to perform this comparison with the same instruments at different latitudes and longitudes (intensity and Dopplergrams). Finally, the third step consists of the extension over the solar cycle with SDO data for the full Sun component.

| Instrument | Duration (h) | Time step (s) | FOV (arc sec) | Pixel sampling (arc sec) | Contrast (original data) | Contrast (deconvolved data) |
|---|---|---|---|---|---|---|
| IRIS (2832 Å) | 6 | 20.8 | 56 x 60 | 0.167 | 0.067 | 0.175 |
| HINODE (4504 Å) | 24 | 50.2 | 75 x 74 | 0.11 | 0.129 | not used |
| THEMIS (6500 Å) | 1 | 37.4 | 100 x 87 | 0.06 | 0.025 | 0.084 |
| SDO (6173 Å) | 3 | 45 | 213 x 185 | 0.50 | 0.023 | 0.077 |
| SIMU (Emergent intensity) | 27 | 60.0 | 131 x 131 | 0.13 | 0.156 | not used |

Table 1: Long sequences of independent observations and contrasts (measured as RMS/average intensities) of the data sets. Estimated height of the continuum at $\tau = 1$, IRIS [Pereira et al., 2013], Hinode [Rieutord et al., 2010], THEMIS [Vernazza et al., 1981], SDO [Norton et al., 2006].

| Instrument | CWL (Å) | FWHM (Å) | Optical resolution (") |
|---|---|---|---|
| IRIS | 2832Å | 15 | 0.33 |
| HINODE | 4504 Å | 7 | 0.20 |
| THEMIS | 6500Å | 100 | 0.25[1] |
| SDO | continuum 6173 Å[2] | | 1.0 |

Table 2: Observation characteristics.

In this paper, we investigate the properties of the exploders through the observations of the ground-based THEMIS telescope, IRIS, SDO, and the Hinode satellites, and finally the simulation. Section 2 describes the data selection and reduction. The "manual" analysis of coordinated observations of exploding granules by THEMIS, IRIS, and SDO is presented in section 3. Section 4 is devoted to the "statistical" analysis of exploding granules from independent long temporal sequences. The results and conclusion are given in section 5.

## 2    THEMIS, IRIS, HMI, Hinode, and simulation data

The main goal of our present work is to evaluate the detection and the expansion of exploding granules at several wavelengths, and at different spatial and temporal resolutions from ground-based and space-borne observatories, and to compare them with numerical simulations. We used two complementary approaches to measure the expansion of the explosive granules: one is called manual and the other statistical. The manual method allows us to follow in detail the evolution of a small number of granules observed simultaneously by several instruments. This method is based on watershed segmentation (see Section 4) and requires the adjustment of some parameters for each time sequence in order to get the best contour of the cells. The final step is to label them. The statistical method does not use simultaneous observations, but deals with a larger number of granules. It can be applied directly to the independent sequences provided by different instruments, and to numerical simulation results, allowing us to identify and label a large number of exploding granules automatically by using the Bovelet segmentation (see Section 5). Thus, it is possible to study the various sequences while preserving their spatial resolution and then normalize the results.

The observations analyzed in this work were taken by different instruments on various satellites: the Hinode spacecraft [Kosugi et al., 2007, Tsuneta et al., 2008], the full Sun HMI [Schou et al., 2012] on board the Solar Dynamics Observatory (SDO Pesnell et al. [2012], the Interface Region Imaging Spectrograph (IRIS) [De Pontieu et al., 2014], and the ground-based THEMIS telescope (Observatorio del Teide, Tenerife).

A coordinated campaign of observations was performed in mid-September 2019 (9 to 14) involving the THEMIS (Tenerife), IRIS, and SDO/HMI satellites. We obtained two sequences of simultaneous observations of the same solar region with THEMIS, IRIS, and SDO/HMI on 10 Sept 2019 with moderate seeing, and on 14 Sept 2019 with good seeing with THEMIS and SDO/HMI. On 10 Sept 2019 the common sequences occurred from 12:27 to 12:53 UT and on 14 Sept 2019 from 08:41 to 09:34 UT at disk center. For the statistical study we added the sequence of Hinode observations (29 August 2007) described in Roudier et al. [2016], and the sequence of simulation results presented in Roudier et al. [2019] and obtained with the 3D magneto-convection code [Stein and Nordlund, 1998]. The characteristics of all sequences are summarized in Tables 1 and 2.

The THEMIS imagery instrument was used to observe the solar granulation at the disk center at $\lambda = 6500$Å. On 10 Sept 2019, the moderate seeing (the adaptive optics and image derotator were off) gave us the possiblity to observe together with IRIS and SDO the same field of view (FOV), but we only obtained a few frames from THEMIS allowing comparison to space-borne observations. On 14 Sept 2019, we obtained a time sequence with very good seeing at the disk center, but only in coordination with SDO/HMI. The mean time step for that date is 37.4 seconds. Prior to restoration, all images were pre-treated for dark and flat-field corrections. We then applied speckle imaging restoration software to the THEMIS data in order to compensate for image degradation due to turbulence. Our software relies on classical Knox-Thompson as-sumptions [Knox and Thompson, 1974], and was largely inspired by the KISIP phase iteration algorithm [von der Luehe, 1993]. The quality of such image restoration has been successfully tested against images from space experiments [Wöger and von der Lühe, 2008]. The final resolution of restored THEMIS images (Figure 1) is 0.2 in the best cases, although the quality is not homogeneous across the whole data set. The restored images were derotated, aligned, destretched, and filtered for p-modes in the k-$\omega$ space [Title et al., 1989] (threshold phase velocity = 6 $kms^{-1}$).

---

[1]This is an estimate: images are derived from 100 frames taken at 30 frames/s and restored after selection of best images.

[2]Gaussian fitting of FeI 6173 line profile, continuum intensity.

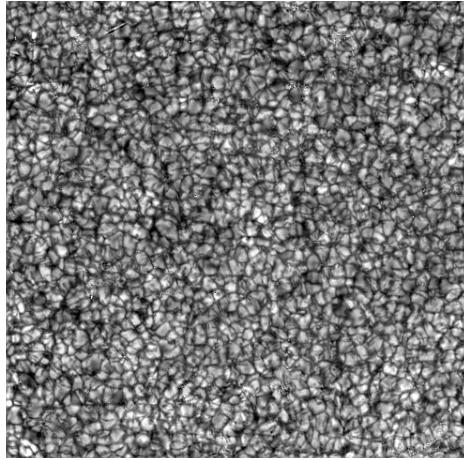

Figure 1: Solar granulation (after image restoration) observed by THEMIS on 14 Sept 2019. The FOV is $60 \times 60$.

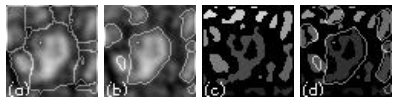

Figure 2: Granules and cell boundaries obtained via various methods. Image (a): Cell boundary obtained by the watershed segmentation algorithm. Image (b): Application of the erosion operator on image (a) gives new cell limits. Image (c): Granule up-flows (bright part) detection by the Bovelet algorithm. Image (d): Superposition of the granule border shown in image (b) and granules detected in image (c).

We used the IRIS slit jaw images to get the temporal evolution of the granulation at $\lambda$ = 2832Å with a time step of 20.8 seconds [Wülser et al., 2018]. By temporal interpolation, we removed the images of the slit on each frame allowing a perfect 2D field. The restored images were then deconvolved by the IRIS transfer function, and finally filtered for p-modes in the k-$\omega$ space (threshold phase velocity = 6 $kms^{-1}$)). The Helioseismic Magnetic Imager (HMI) on board SDO provides uninterrupted observations over the entire disk. This gives a unique opportunity for extracting solar granulation characteristics during a long period of time with uniform observations. We used SDO/HMI data observed simultaneously with THEMIS at the disk center. The original pixel size of images is close to 0.5. Each image was deconvolved by the HMI transfer function [Couvidat et al., 2016] and rebined by a factor of two (pixel of 0.25). This last operation allows an increase in the number of pixels for each granule and allows better temporal tracking.

We also used data sets from the Solar Optical Telescope (SOT) on board the Hinode mission. For our study we used blue continuum observations, on 29 August 2007, at $\lambda$ = 4504Å from the Hinode/SOTBFI (Broadband Filter Imager). After the flat-field and dark corrections [Lites and Ichimoto, 2013], the images were aligned and filtered for p-modes in the k-$\omega$ space (threshold phase velocity = 6 $kms^{-1}$). It should be noted that, for each data set considered, the conversion from arcsecond to kilometer takes into account the distance between the corresponding instrument and the Sun.

The magnetohydrodynamics simulation provided a 27-hour duration sequence of the (pseudo white light) emergent intensity, surface velocity (Vx,Vy,Vz). For this sequence, we used only intensity and velocity vectors at $\tau$ = 1 (z = 0 Mm) (Roudier et al. [2019]. The pixel size is 96 km.

## 3   Image segmentation

The measurement of the photosphere properties requires a definition of a physical entity that best characterizes solar granulation. Two structure types can help to describes solar granules. One of these entities is called a cell and contains both a rising (bright granule, upflow) and a falling portion (dark intergranule, downflow) of the plasma.

The other entity of interest is associated with rising plasma (bright granules only, upflow) in the case of the solar turbulent convection with bright elements visible in the intensity field called granules. The detection of cells can be carried out by various methods, such as the iterative procedure Medial Axis Transform (MAT) used by Berrilli et al. [2002]. We used the watershed method for cell detection. In this approach, the topographic surface is mainly built from an image intensity gradient, in both directions, x and y, since the object edges (i.e., watershed lines) are located at pixels with high gradient values [Derivaux et al., 2010]. The watershed delineation algorithm used here provides a complete partition of watersheds, and every pixel in the image belongs to one region (cell). The watershed method allows us to detect the granule and part of the surrounding intergranular area, taking into account the part of the granule up to its neighbors. The left part of Figure 2 (a) shows the cell delineated by the watershed method, applied to IRIS data (i.e., Figure 5 top) which are similar to those obtained by Berrilli et al. [2002]. However, we applied an erosion operator to the detected structure in order to avoid labeling artifacts in the temporal label of each granule. The border of the detected granule resulting from that erosion is shown Figure 2 (b). These structures, containing bright granules (upflow) and some dark intergranules around (downflow), are called "cells" in the following.

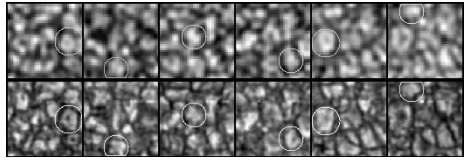

Figure 3: Six exploding granules detected (circle) during 1164-second sequence over a field of $10 \times 10$ on SDO (top row) and IRIS (bottom row) data with the same pixel sampling. Images from left to right correspond to detection times at 0, 270, 624, 811, 1019, and 1164 seconds.

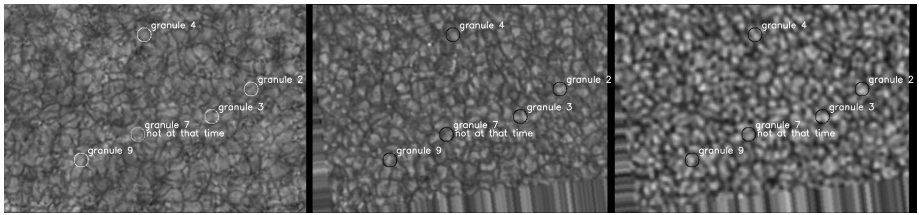

Figure 4: Location of the five granules observed by THEMIS (left), IRIS (middle), and SDO (right) on 10 Sept 2019. The FOV is $56 \times 38$.

To isolate granule upflows (bright part of the granule), we used the Bovelet segmentation described in Bovelet and Wiehr [2001]. This pattern recognition algorithm, called the multiple-level tracking (MLT) algorithm, is based on multiple intensity clips. This procedure combines intensity clip, recognition, and subsequent extension of individual shapes on multiple intensity levels. Thus, shapes recognized and extended via various intensity levels, continue to fill their final contours of brightnesses observed. Adjacent shapes are kept separated by an optional minimum distance up to two pixels [Bovelet and Wiehr, 2001]. An example of the granule detection with the segmentation algorithm applied to the same granule as above is shown in Figure 2 (c).

Figure 2 (d) shows the comparison between the two segmentations used in this paper, where the areas are not different in size. This is not the case when the shapes of the granules are more irregular.

# 4 Manual analysis of exploding granules of simultaneous observations and simulation

In the manual approach each granule area is extracted by using the watershed method. It requires adjusting some parameters, like the size of the opening morphological disk and intensity threshold, for each sequence to get the best cell contour of the granule (to the eye). The watershed method allows us to define the cell of the granule, taking into account the part of the granule up to its neighbors. This step cannot be automated and homogenized for all sequences, which is why we only selected a small number of exploding granules for each sequence (five on 10 Sept and ten on 14 Sept). After the segmentation the final step is to label each granule in space and time. In order to compare the images from the different telescopes, we chose to interpolate the data to the pixel size of THEMIS. The granule area error bar is mainly related to the pixelation and segmentation process. We estimated the radius error measurement to be one pixel. This gives, in the worst cases, an estimate of 0.3 $kms^{-1}$ on the expansion velocity error at time t=600 seconds from the first image.

To control the exploding granule detection in IRIS and SDO observations, we used IRIS data with the highest spatial resolution as reference. Figure 3 confirms that exploding granules are visible in both data sets, and reveals the high quality of the deconvolved SDO data. The time evolution of both sequences confirms the good detection of exploding granules. Six exploding granules are thus detected in both observations during 1164 seconds sequence over a FOV of $10 \times 10$ (1 arcsec = 729.6 km) giving an occurrence rate of exploding granules of $8.9 \pm 0.710^{-11}$ $km^{-2}s^{-1}$, quite similar to those given by Palacios et al. [2012] of $7.710^{-11}$ $km^{-2}s^{-1}$. The last data were acquired with the Sunrise balloon-borne mission (instrument IMaX; Martínez Pillet et al. 2011), close to spatial conditions without the perturbations induced by the Earths atmosphere, on a short time sequence (around 20 minutes) and small FOV ($45 \times 45$). The similarity of the occurrence rate of exploding granules between the data obtained with the SDO moderate spatial resolution of 0.5and the high resolution images from IMaX of 0.15 reinforces the use of the deconvolved images from SDO and rebinned by a factor of two (pixel of 0.25) to study exploding granules on large FOVs over the solar surface and during long periods of time.

## 4.1 Local exploding granules observed by THEMIS, IRIS, and SDO on 10 Sept 2019

Taking into account the moderate seeing at THEMIS on 10 Sept 2019, we selected only five granules over 1500 s (Figure 4) . Those granules in our selection have a starting area between 1 and 1.5 $Mm^2$, which corresponds to a radius of 700 km assuming a disk shape for the granule area, and grow up to 3.5 $Mm^2$ (radius of 1050 km) in the first 600 s. An expansion velocity is computed from the difference in radius between the first image and the images at times 60 and 600 seconds. The expansion velocity of the five selected granules is found to be around 1 $kms^{-1}$ in the first minute and decreases to 0.45 $kms^{-1}$ after 600 seconds. Figure 5 gives an example of the evolution of such a granule observed by the three instruments. We find a good agreement in the area evolution of that granule (shown in Figure 6) computed from the three instruments at different spatial resolutions. The time derivative of the area provides an estimation of the expansion velocity, but is too noisy to be plotted and to be conclusive. The correspondence of the area evolution between the three instruments is available for the five selected granules. Figures 7 and 8 give an example the 3D evolution (x,y,t) of the same granule between 0 and 1500 s showing a great similarity between segmentations applied to IRIS and SDO data; SDO areas appear more noisy due to the lower spatial and temporal resolutions. Figure 9 confirms the very good correlation (R=0.92) of the measured radius of the granule on both IRIS and SDO data. Although the SDO data have a lower spatial resolution (by a factor of 2), the combined effects of the oversampling just after the deconvolution and the segmentation processing, which is sensitive to the intensity gradients, provide area measures slightly larger than IRIS when all images are interpolated to the THEMIS pixel size. It can be seen in Figure 5 where the segmented granule areas of SDO and IRIS are shown. Due to the lower resolution the radius measured with SDO is slightly larger than that measured with IRIS.

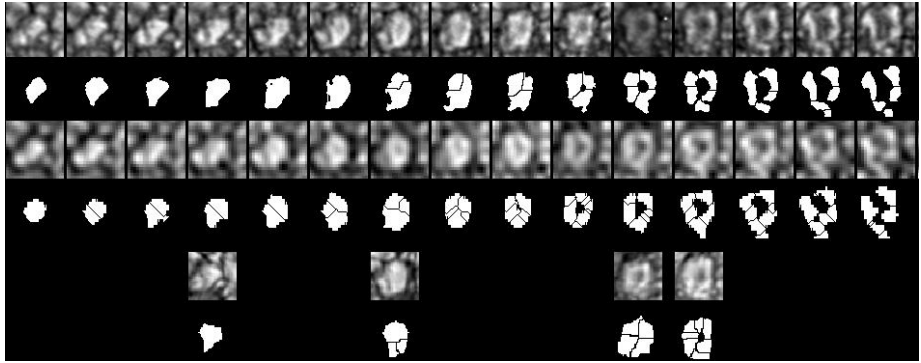

Figure 5: Evolution during the first 416 s of an exploding granule IRIS (top), SDO (middle), and THEMIS (bottom) on 10 Sept 2019. The FOV is $3.6Mm \times 3.6Mm$ and the time step is 45 seconds between images.

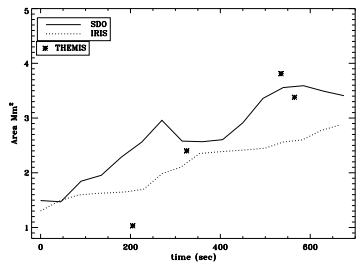

Figure 6: Area evolution of the same exploding granule in Figure 5, from IRIS, SDO, and THEMIS on 10 Sept 2019.

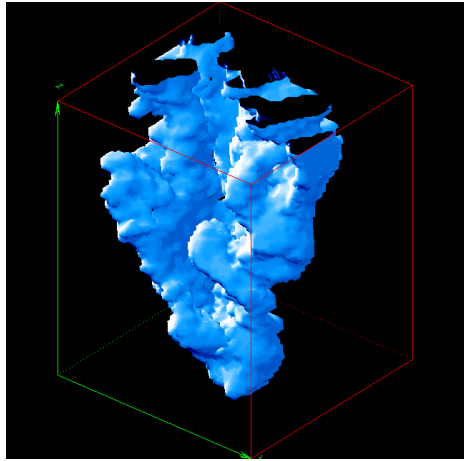

Figure 7: Three-dimensional evolution (x,y,t) of the same exploding granule in Figure 5, from IRIS on 10 Sept 2019: X = 3.6 Mm, Y = 3.6 Mm, duration= 1487 s.

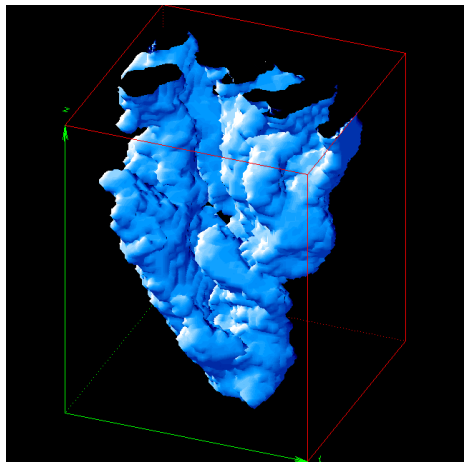

Figure 8: Three-dimensional evolution (x,y,t) of the same exploding granule in Figure 5, from SDO on 10 Sept 2019, granule 4: X = 3.6 Mm, Y = 3.6 Mm, duration = 1485 s

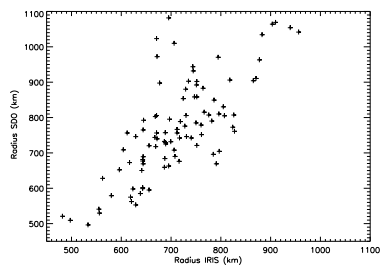

Figure 9: Plot of the radius of all granules measured by IRIS and SDO data on 10 Sept 2019.

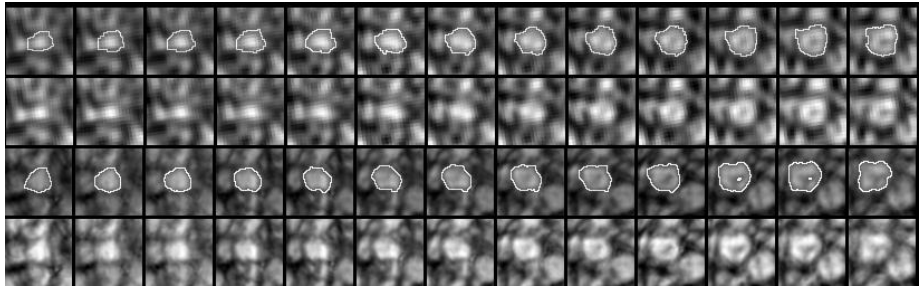

Figure 10: Evolution of the exploding granule from SDO (top lines 1, 2) and THEMIS (bottom lines 3, 4) on 14 Sept 2019. The FOV is $5.1Mm \times 5.1Mm$ and the time step is 45 seconds between images.

## 4.2 Local exploding granules observed by THEMIS and SDO on 14 Sept 2019

On 14 Sept 2019 we selected ten exploding granules from the 53 min sequence at the disk center from THEMIS and SDO simultaneous observations. In that sequence the starting area we selected is between 2 and 3.0 $Mm^2$ (radius 800 and 980 km) and grows to 5.5 $Mm^2$ (radius 1260 km) in the first 600 s. The larger area on that date relative to the 10 Sept 2019 measurement for the THEMIS data is due to the higher seeing quality. The segmentation parameters of SDO are fixed, for all sequences, to get the best granule areas close to the THEMIS values. The expansion velocity, however, is found to be approximately the same as on 10 Sept 2019. It is around 1 $kms^{-1}$ in the first minute and decreases to 0.5 $kms^{-1}$ at 600 s. Figure 10 gives an example of the evolution of such a granule, observed by THEMIS and SDO. Figure 11 suggests that the granule area A(t) is close to the analytical form A(t) = a + b t, where a and b are two constants. This behavior is corroborated by statistical results in the next section. In this case, the equivalent radius R(t) of granules such that A(t) = pi * $R^2(t)$ provides the expansion velocity v(t) = dR(t)/dt that decreases with R(t). The area evolution of that granule is shown in Figure 11 with a good correlation between the two instruments at different spatial resolutions, in agreement with results shown in Figure 6. The plot of the granule radius and the estimated expansion velocity (at t = 120 s) relative to the area of first image is shown in Figure 12. We note a moderate correlation (R=-0.54) between the size of the granule and expansion velocity. This indicates a trend of higher expansion velocities for smaller granules; in other words, the expansion velocity of granules decreases with time.

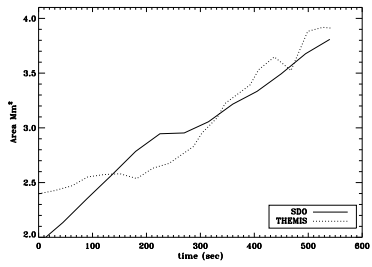

Figure 11: Area evolution of the exploding granule from SDO and THEMIS on 14 Sept 2019.

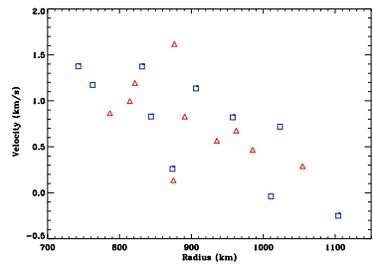

Figure 12: Estimated expansion velocity vs. granule radius at t = 120 s from THEMIS (blue square) and SDO (red triangle) data on 14 Sept 2019.

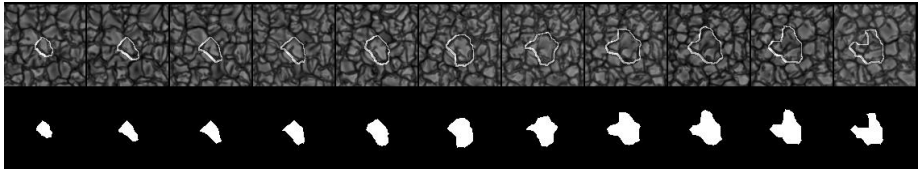

Figure 13: Example of an exploding granule evolution from the simulation. The temporal step between presented images is 120 s. The FOV is $9.7 Mm \times 9.7 Mm$.

### 4.3 Exploding granules detected in the simulation and interpretation of results

In the simulation, we selected nine exploding granules. In the same way as for observations, the granule area is extracted by using the watershed method. The starting area of the granules is between 1 and 3.5 $Mm^2$ (radius 1050 km) and grows up to 7 $Mm^2$ (radius 1500 km) in the first 600 s. The final sizes are larger than in the observation (14 Sept 2019, see above). In all cases the area grows approximately linearly during their lifetimes. Surprisingly, we see that the module of horizontal velocities inside the granule is quite constant during the growth phase in 61% of our simulation results, even though the granule area expands regularly (see Figure 13). One explanation lies in the repartition of the horizontal velocities in the simulation, which are located preferentially on the borders of the granules with lower velocities in the center. This was observed previously by Oba et al. [2020] and Oba [2018]. The mean expansion velocity at t = 660 s of our sampling is 0.82 $kms^{-1}$, which is higher than found in the observations. Figure 13 shows the evolution of one of the selected exploding granules and its segmented area. The area evolution of that granule is shown in Figure 14. The expansion velocity decreases with increasing granule size, as for the observations (not shown).

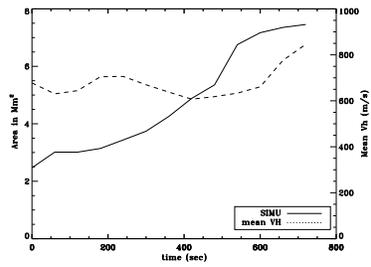

Figure 14: Area evolution of the exploding granule and of the module of the horizontal velocity (Vh), provided by the simulation, inside the granule area.

# 5 Statistical analysis of exploding granules from long sequences of observations and simulation

For the statistical approach, each granule area is extracted by using the Bovelet segmentation [Bovelet and Wiehr, 2001]; the final step is to label them. In comparison with the watershed segmentation (cell detection), the Bovelet segmentation isolates only the bright parts of the granules (upflows). The area measurement error on each granule, which is linked (as in the manual method) to pixelation, and the segmentation process is on the same order giving an error bar of 0.3 $kms^{-1}$, in the worst cases, on the measured expansion velocity at time t=600 seconds from the first image.

## 5.1 Segmentation of temporal sequences

The manual method allows us to compare the same granule observed by THEMIS, SDO, or IRIS and to plot their areas as a function of time, but not the expansion rate, which is a noisy derivative. Conversely, the automatic method has the advantage of working on tens or even hundreds of explosive granules, giving the evolution of an average area and mean expansion velocity that can be calculated using statistics.

Each sequence was segmented and granules were labeled as described in [Roudier et al., 2003]. Granules form Trees of Fragmenting Granules (TFGs or more simply families of granules). Families originate from a single parent exploding granule and their development (often lasting several hours) can be followed as a function of time. The segmentation was done for all sequences with the same parameters. The filling factor of the granulation, representing the relative value of a granular area in comparison with a total area, was fixed to 0.4 for all sequences. This threshold value allows us to get the bright part (upflow) of the granules as shown in Figure 2 image (c).

The emergent intensity of the simulation corresponds to z = 0 km of the atmospheric model used by the code. After segmentation, the families were sorted by decreasing size (i.e., the time integrated total surface over the sequence). Families already existing at time t = 0 were eliminated. We considered only families forming during the sequence, in order to study the birth and growth of the "parent" granule and its consecutive explosion forming several "children" granules, together with their development. The "birth" time is defined as the time when the segmentation allows the detection of a new granule that is at the origin of a large family. Hence, the method provides statistics of the formation and expansion of parent granules at the origin of the largest families, in terms of area (Figure 15) and expansion speed (Figure 16). The results suggest that the mean initial expansion speed is in the range $2 - 3$ $kms^{-1}$ for the multi-wavelength observations and simulation, and decreases in a few minutes to about 1 $kms^{-1}$ or less. The mean area of parent granules exhibits a similar behavior both for observations and simulation. It should be noted, however, that sequences are much shorter for THEMIS (Table 1). Sequences also have different wavelengths, bandpass, optical resolutions (Table 2), and contrasts (Table 1). The IRIS continuum forms above the other observations (Table 1).

The area growth rate is higher in THEMIS data probably due to the the phase diversity processing. We recall that each THEMIS image comes from the combination of

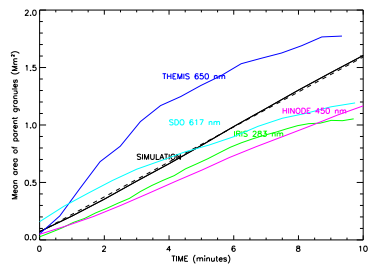

Figure 15: Mean area of the parent granules leading to the development of the largest families. The initial area is about $0.05 Mm^2$. The statistical method allows us to detect the birth of a series of exploding granules, which later form large TFGs. The dashed line corresponds to an area law of the form $A(t) = a + b \times t$, where a and b are constants, fitted to simulation results.

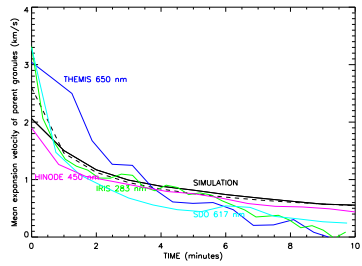

Figure 16: Mean expansion velocity of the parent granules leading to the development of the largest TFGs. The dashed line corresponds to the expansion velocity dR/dt, where R(t) is the radius of the disk area such that $\pi \times R^2(t) = A(t) = a + b \times t$, where a and b are constants, fitted to the simulation.

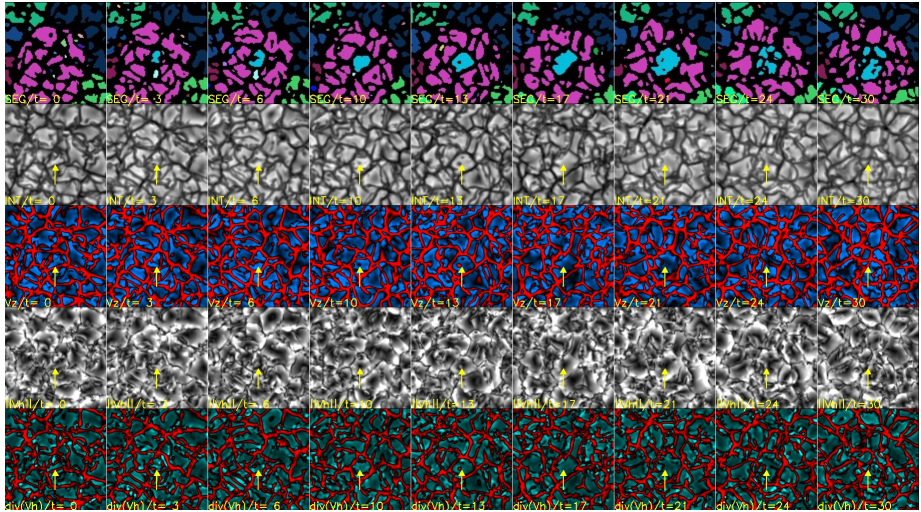

Figure 17: Typical parent granule issued from the numerical simulation, at the origin of the formation of a large TFG. From left to right: time t = 0 (birth), 3, 6, 10 (growing phase), 13, 17, 21, 24 (explosion), and 30 minutes. The FOV is $9.6 Mm \times 9.6 Mm$.

Line 1: intensity segmentation (the exploding granule is shown in blue);

Line 2: emerging intensity (in gray);

Line 3: vertical velocity: redshifts (downflow) and blueshifts (upflow);

Line 4: horizontal velocity module (in gray);

Line 5: horizontal velocity divergence (red for converging flow, blue for diverging flow).

The arrows indicate the location of the exploding granule.

typically 50 frames obtained under varying seeing conditions; the restoration process enhances the contrast by a factor of 3, but also has the tendency to spread and deform structures. Hence, the expansion velocity, as a derivative of the area, is affected by the restoration. This non-isotropic spread appears to produce larger granules with faster expansion. We think that the area growth rate for THEMIS data is due to the seeing effects and to our reduction processing. Data issued from free seeing space-borne instruments show a slower area expansion. We do not observe any relationship between area expansion and the relative altitude of observations.

As parent granules are detected earlier than can be done with the manual method described in the previous section, the two methods provide complementary results, at different development stages. The characteristics of expansion velocities are summarized in Table 3.

## 5.2 Predictions of the numerical simulation

Previous work about exploding granules was performed on a small number of granules because precise measurements are delicate, and time sequences at high resolution in intensity or Doppler representations are generally short. We are able, for the first time to our knowledge, to get the temporal evolution of the mean expansion velocity from our statistical analysis. We analyzed 637 exploding granules at the origin of the largest families of granules of the 27-hour sequence. In Figure 16, we observe the same trend and close amplitude of the temporal decrease of expansion velocity in the simulation as we measure from observations. The advantage of the simulation is that it allows us to map the temporal evolution of the vertical velocity with depth below exploding granules.

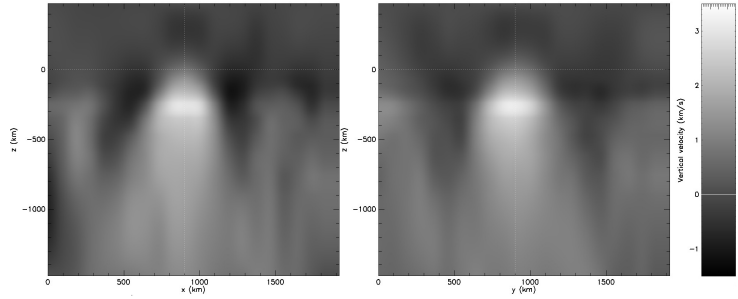

Figure 18: Cuts in the planes (x,z) (left) and (y,z) (right) of the velocity Vz averaged over 48 parents of exploding granules at their birth at z=0 km. The upflow (plus) and downflow (minus) are white and black, respectively. On the ordinate axis, the altitude z is positive above z=0 and negative below (toward the convection zone).

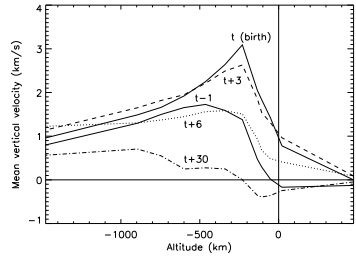

Figure 19: Temporal evolution of the average vertical velocity in depth, from -1500 km to +500 km, the surface being at 0 km, computed from 48 exploding granules at their center. On the surface Vz is at 0.9 km/s, but there is a peak at 3 km/s at 200 km below the surface at the birth of the granule. The different curves correspond first to the onset of granule parent at t(birth)-1 minute up to t(birth)+30 minutes. We observe a clear increase in the upflow velocity just before the birth of the exploding granule at z=0 (surface) and then a monotonic decrease in time of the Vz component.

| Instrument | Number of parent granules | Mean initial expansion speed (km/s) | Mean expansion speed after 3 min. (km/s) | Standard deviation $\sigma$ (km/s) |
|---|---|---|---|---|
| IRIS 2832Å | 166 | 3.30 | 1.0 | 1.0 |
| HINODE 4504 Å | 381 | 1.9 | 0.90 | 0.5 |
| THEMIS 6500Å | 131 | 3.05 | 1.2 | 1.6 |
| SDO 6173Å | 161 | 3.30 | 0.7 | 0.8 |
| SIMULATION | 637 | 2.05 | 0.9 | 0.6 |

Table 3: Mean initial expansion speed and mean expansion speed after 3 min for observations and simulation

The initial expansion speed of parent granules is in the range $1 - 3$ $kms^{-1}$ with a 2.0 $kms^{-1}$ mean value. The average expansion speed decreases from 2.0 $to$ 1.0 $kms^{-1}$ in only 3 minutes (or from 2.0 $to$ 0.5 $kms^{-1}$ in only 10 minutes). The simulation shows that it is not the case, however, for the horizontal plasma velocity, which remains approximately constant (1.5 $kms^{-1}$). The simulation also predicts that the divergence of initial horizontal velocities is always positive (outward-directed) and lies in the range 0.008 $s^{-1}$ to 0.013 $s^{-1}$ (0.01 $s^{-1}$ mean value). The vertical plasma velocity is upward-directed (1.5 $kms^{-1}$ mean value) and does not vary much.

Examples from the simulation show that horizontal velocities in granules increase smoothly from the center to the border, and disappear above intergranular lanes (where the velocity vector turns downward). The plasma diverges inside granules and converges above the intergranular lanes. Some examples show that the explosion of granules starts with the formation of narrow dark points, exhibiting downward motion and converging horizontal flow. The dark points then extend, split the granule, and form new intergranular lanes (Figure 17).

Figures 18 and 19 describe the evolution of the granules and intergranular lanes shown by the simulation. Figure 18 shows cuts in the (x,z) and (y,z) planes, of vertical velocity Vz, averaged over 48 exploding parent granules. Figure 19 is a plot of the mean vertical velocity temporal evolution, computed from the 48 parent granules at their center, from -1500 km to +500 km (with 0 km being the surface). At an average granules birth (fragmentation from a parent), Vz at the surface is 0.9 $kms^{-1}$, but a peak of 3 $kms^{-1}$ at 200 km below the surface can be seen in Figure 19. The simulation shows a clear increase in the upflow velocity at all depths between t(birth)-1min and t(birth) of an average exploding granule. From z=0 (surface) down to -200 km, a monotonic decrease in time of the Vz upflow component is observed. This temporal upflow evo-lution appears in Figure 18 as a rising "bubble" of plasma that emerges (white), with descending movements (black) on either side. We note a maximum upflow around 200 km below the surface (brightest white) and clear down flow (darkest black) down to -700 km around an average exploding granule.

At the top of the photosphere, 300 km above the emergence, the speed is downward. The lower Vz velocity in time may be responsible for insufficient pressure to move mass laterally [Nordlund et al., 2009]. According to these authors "the density builds up over the granule until the pressure is raised sufficiently to expel it. The excess pressure also decelerates the upflow and thus reduces the energy transport to the surface, in particular near the granule center, which then cools. Hence, as granules grow, the upflow velocity near their center decreases. The continuous decrease in expansion velocity during the fragmentation process, for all observations, is compatible with their physical description from the simulation. The evolution of the surrounding granules also contributes to compelling the expansion of the granule. The combination of the two phenomena, the decrease in the Vz amplitude in time and close proximity of granules, can explain the mean expansion velocity decrease that we observe.

# 6   Results and conclusion

From ground-based instruments (THEMIS) and space-borne instruments (IRIS, SDO/HMI, Hinode) the evolution of explosive granules is studied at different wavelengths and various spatial resolutions. The analysis of the temporal sequences was performed using two methods: manual and statistical.

The manual method allows us to follow individually the exploders observed simultaneously by the different instruments. In this method the watershed segmentation is applied to extract the cell (bright and intergranular parts) of each granule. This approach requires adjusting the segmentation parameters for each sequence limiting the number of exploders studied. Due to the small sample of selected exploders, we only followed the granule area evolution, the derivative being too noisy. The simultaneous temporal evolution of the exploders' area with the THEMIS, IRIS, and SDO instruments are found to be well correlated. An exploder's area evolves between 2 and 7 $Mm^2$ in the first 600 s giving an average velocity of the order of 1.1 $kms^{-1}$ in agreement with previous works [Hirzberger et al., 1999, Title et al., 1989, Namba, 1986]. The expansion rates are higher in the development phase when the granules are the smallest in both observations and simulation.

The statistical method allows access to a large number of exploders with a fixed segmentation based on Bovelet segmentation with a granule filling factor fixed to 0.4 for all sequences. Only the bright parts of the granules are extracted with this technique. All the dynamical properties found on a large number of exploders confirms the previous results found by the manual method. The large number of analyzed exploders allows us to derive the expansion velocity. These velocities decrease rapidly in the first two minutes.

The temporal evolution of the mean expansion velocity that we observe is compatible with the physical granule evolution described in Nordlund et al. [2009], where large pressure perturbations drive their horizontal flows. Both the excess pressure at the center of the exploding granule, which decelerates the upflow and then reduces the energy transport to the surface, and the evolution of horizontal flow of neighbor granules, contribute to reduce the expansion velocity of the exploding granule. We observe that the temporal upflow evolution looks like a plasma rising bubble that emerges, with descending movements on either side.

The differences in evolution of the granule's area as a function of wavelength does not seem related to the mean altitude of the observations. The seeing correction is the main reason, in the case of the THEMIS observation, for larger growing rates. In addition, THEMIS temporal sequences are short, justifying new campaigns with adaptive optics in order to avoid successive frame combinations. The very good spatial resolution of the Hinode, IRIS, and simulation data leads to close results which could serve as reference.

Our study reveals the great potential of the SDO/HMI deconvolved observations, showing an occurrence rate of exploding granules similar to that in the preceding observations [Palacios et al., 2012], to detect and measure the dynamic of explosive granules on the full Sun surface. However, moderate spatial resolution of these data leads to a slight overestimation of the evolution of areas and expansion rates, regardless of the method used.

This work was granted access to the HPC resources of CALMIP under the allocation 2011-[P1115]. Thanks to SDO/HMI , IRIS , and Hinode/SOT teams. The THEMIS telescope is supported by the French Centre National de la Recherche Scientifique (CNRS - INSU). We thank all the team of THEMIS for adjusting the telescope during our observing campaign. All of the authors would like to thank R. Stein for the use of the 96 Mm x 96 Mm x 20 Mm simulation.We thank the anonymous referee for his/her careful reading of our manuscript and his/her many insightful comments and suggestions.